# Is Computation Reversible?

Michael C. Parker* & Stuart D. Walker**

*Fujitsu Laboratories of Europe Ltd., Columba House, Adastral Park, Ipswich IP5 3RE, UK*

**Department of Electronic Systems Engineering, University of Essex, Wivenhoe Park, Colchester CO4 3SQ, UK*

**Recent investigations into the physical nature of information and fundamental limits to information transmission have revealed questions such as the possibility of superluminal data transfer [1] or not [2]; and whether reversible computation (information processing) is feasible [3]. In some respects these uncertainties stem from the determination of whether information is inherent in points of non-analyticity (discontinuities) [4] or smoother functions [5-7]. The close relationship between information and entropy is also well known [8,9], e.g. Brillouin's concept of negentropy (negative entropy) as a measure for information [10]. Since the leading edge of a step-discontinuity propagates in any dispersive medium at the speed of light in vacuum as a precursor to the main body of the dispersed pulse [11], we propose in this paper to treat information as being intrinsic to points of non-analyticity (discontinuities). This allows us to construct a theory addressing these dilemmas in a fashion consistent with causality, and the fundamental laws of thermodynamics. A consequence of our proposition is that the movement of information is always associated with the dissipation of heat, and therefore that the concept of reversible classical computation is not tenable.**

Consideration of information as being intrinsic to points of non-analyticity is akin to the treatment of it as a 'particle', despite it having apparent dual "wave-particle" (i.e. localised and/or distributed) characteristics [4-7]. Such complementary aspects are of course paradoxical in nature, but by adopting this particular viewpoint, the following analysis aims to be self-consistent. Landauer's principle [8,12] in its 'static' form states that erasure of information requires energy and hence is associated with an increase in entropy; whereas the creation of information doesn't require energy and so is not associated with an increase in entropy. We have recently formulated a dynamic version of this principle [13], where transfer of information from A to B can be thought of as the annihilation of the information at A (therefore accompanied by an increase in entropy) followed by its re-creation at B (which is not associated with any change in entropy). It is well known that computation is associated with the 'shuttling' of input information through logic gates according to a given algorithm (the program), in order to transform that information into an alternative 'more interesting' output form. A computation does not intrinsically add 'new' information to the input information; and the use of 'reversible' logic gates, such as Toffoli gates to ensure zero information loss, means that a computation doesn't necessarily lose any information either [3,14]. In conjunction with the static form of Landauer's principle, this has led to the concept of 'reversible' computation, with its associations with reversible thermodynamics, and the implication of energy-neutral computation.

To continue our discussion of information transfer in a meaningful sense, the concept of differential information must be introduced [13,15], since spatial differences in information are best described by the differential information. We note, that in the infinitesimal limit of the well-known discrete summation describing information, there is a diverging part that we ignore when discussing information transfer. This is because when considering differences between information at different spatial positions, the diverging parts cancel; the constant (infinite) divergent part due to the infinitesimal limit being the same everywhere. We have previously shown that the differential information *I*, i.e. that information which can be transferred from one spatial location to another, is given by the sum of the associated residues, $2\pi i \sum R$, from the Cauchy residue theorem [13], where $\psi(x)$ is the normalised wave function encoding the information:

$$I = \int_{-\infty}^{\infty} |\psi(x)|^2 \log_2 |\psi(x)|^2 \, dx = 2\pi i \sum R \quad \text{(1a)} \quad \text{and} \quad \int_{-\infty}^{\infty} |\psi(x)|^2 \, dx = 1. \quad \text{(1b)}$$

Residues are a consequence of localised discontinuities or points of non-analyticity in the complex plane, and as indicated in Figures 1a and 1b these points can be considered to be the inherent 'location' of the differential information. Functions obeying the Cauchy-Riemann equations allow the process of analytic continuation, such that the function can be completely reconstructed in a self-consistent fashion from any point in those regions where the Cauchy-Riemann symmetry conditions hold [16]. In regions where the Cauchy-Riemann conditions do not hold, analytic continuation of the function cannot be performed and the function becomes in effect "non-predictable" in those regions. Figure 1b shows such a discontinuity due to a simple pole in the complex plane at $z_0$. We note that a simple pole obeys the Paley-Wiener criterion for causality [17], as well as being square-integrable and having zero power at infinite frequencies. The lack of smoothness at the peak of the function is obvious, as is the case for any discontinuity. Points of non-analyticity are therefore inimical to any assumptions of 'smoothness', such that in general, when the system is allowed to dynamically evolve in time, assumptions of adiabiticity are not tenable, and one would expect the entropy to increase.

Present day photonic data links may be taken as an example of this principle of entropy increase with information transfer. Here, light modulated with data is transmitted through a passive medium, e.g. in free-space, or down an optical fibre. The light will either suffer the degradations of diffraction in the former case, or waveguide and material dispersion in the latter. Both these processes lead to the gradual attenuation of the light as it travels ever-longer distances, and a reduction in optical signal to noise ratio (SNR). Shannon's theorem [18] requires that the channel capacity must therefore also reduce with distance, e.g. [19], given no redundancy in the information for error-correction purposes. The reduction in capacity reveals itself as an increase in the associated bit-error-rate (BER), and consequent loss of information. Assuming information and entropy to be negatively correlated [10], this means an increase in entropy. The use of amplifiers to compensate for such attenuation also fundamentally leads to an increase in entropy, due to the impossibility of noiseless amplification [20] (noise must always be added to an amplified signal), and the no-cloning theorem at quantum levels [21]. At the receiver end, imperfect photon to charge carrier conversion intrinsically degrades signal strength, and other well-known effects such as photon related shot noise, leakage current and noise from active and passive components [22], all act to further degrade electrical SNR. Even coherent detection schemes suffer impairments as the photon statistics of the local oscillator also serve to introduce noise. Hence data transmission is associated with an increase in entropy. In figure 2, we attempt to schematically show a number of different aspects of the information transmission problem. At a time $t_0$, a localised quantity of information $I(z_A)$ moves in the direction from A towards B. Erasing that information at its spatial location A requires energy $\Delta E$. Arriving at B at a later time $t_1$, Landauer's principle determines that no energy is required to reinstate (create) that information $I$, now at location $z_B$. Overall, assuming the set-up can be characterised by a finite temperature $T$, the entropy of the system increases by $\Delta S = \Delta E / T$. The required amount of energy $\Delta E$ tends to increase with distance, for example, for the case of photonic networks energy is dissipated due to attenuation at typically 0.2dB/km [23].

Sommerfeld and Brillouin studied the propagation of a step-discontinuous pulse through a dispersive (causal) medium [11], and found that the pulse leading edge propagates at the speed of light in vacuum *c*, via forerunners as a precursor to the main body of the dispersed pulse, whatever the refractive index of the medium. Since the point of information must essentially remain a discontinuity, it therefore travels at *c*. This is therefore consistent with classical causality (where superluminal data transfer is not allowed), as well as agreeing with the spirit of Sommerfeld and Brillouin's work, where the forerunners that potentially could carry a signal, propagate at *c*. Hence, the relation $L/\Delta t = |z_B - z_A|/(t_1 - t_0) = c$ holds. We note, that the impossibility of superluminal information transmission is, in itself, an indication of the discontinuous nature of information.

Moore's Law implies that logic gate density will continue to double every 18 months, such that computers will continue to decrease in size for a given processing power [15]. Ultimate physical limits to computers have already been discussed in detail [14], such that computers will always have a finite volume, i.e. any physical process (such as a computation) cannot take place in a point volume. A finite volume computer means that information will always have to travel some distance whilst being shuttled through the various logic gates, such that heat will always be dissipated due to the dynamic form of Landauer's principle. Hence, although for a given processing power computers will dissipate less heat as they reduce in size, they will only become 100% efficient (i.e. dissipate no heat) in the limit of being infinitely small. We note that quantum computing [24] appears to suffer a similar restriction, since the quantum system needs to be completely isolated from the environment to avoid decoherence of the wave function. Entropy might then be expected to remain constant during computation. However, control of the environmental isolation becomes increasingly difficult as the physical size of the quantum computer increases [25].

Overall, our investigations indicate that a perpetual calculating machine is not possible, with computers operating under the same constraints as conventional mechanical machines: obeying the Second Law of Thermodynamics and always operating at below 100% efficiency. We conclude that a finite physical classical computer will always have to dissipate some heat due to the movement of information, such that the concept of thermodynamically reversible computation is not tenable.

Figure 1: a) A holomorphic function f(z) allowing complete analytic continuation across the complex z-plane such that it contains zero differential information, b) A meromorphic function with a simple pole at $z_0$ in the z-plane, with a finite residue sum, containing differential information that can be transferred.

Figure 2: Transfer of information from A to B requires energy $\Delta E$ to 'erase' the information *I* at position $z_A$, time $t_0$, followed by recreation of the information at position $z_B$, at a later time $t_1$, such that the speed of information transfer is $|z_B - z_A|/(t_1 - t_0) = c$. Given a system temperature of *T*, the overall entropy will increase by an amount $\Delta E / T$.

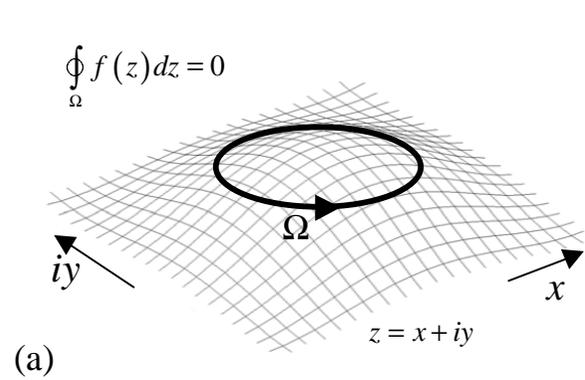 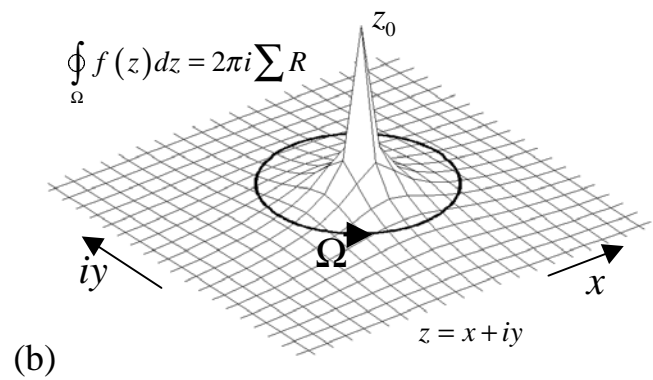

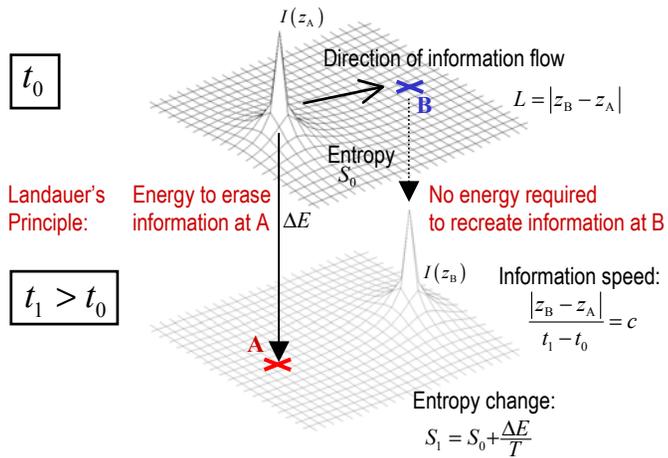